# Using RuleBuilder to graphically define and visualize BioNetGen-language patterns and reaction rules

Ryan Suderman, G. Matthew Fricke, and William S. Hlavacek

RuleBuilder

## Affiliations

Ryan Suderman and William S. Hlavacek

*Theoretical Biology and Biophysics Group, Theoretical Division and Center for Nonlinear Studies, Los Alamos National Laboratory, Los Alamos, NM 87545;*

G. Matthew Fricke

*Department of Computer Science, University of New Mexico, Albuquerque, NM 87131;*

# Summary

RuleBuilder is a tool for drawing graphs that can be represented by the BioNetGen language (BNGL), which is used to formulate mathematical, rule-based models of biochemical systems. BNGL provides an intuitive plain-text, or *string*, representation of such systems, which is based on a graphical formalism. Reactions are defined in terms of graph-rewriting rules that specify the necessary intrinsic properties of the reactants, a transformation, and a rate law. Rules may also contain contextual constraints that restrict application of the rule. In some cases, the specification of contextual constraints can be verbose, making a rule difficult to read. RuleBuilder is designed to ease the task of reading and writing individual reaction rules, as well as individual BNGL patterns similar to those found in rules. The software assists in the reading of existing models by converting BNGL strings of interest into a graph-based representation composed of nodes and edges. RuleBuilder also enables the user to construct de novo a visual representation of BNGL strings using drawing tools available in its interface. As objects are added to the drawing canvas, the corresponding BNGL string is generated on the fly, and objects are similarly drawn on the fly as BNGL strings are entered into the application. RuleBuilder thus facilitates construction and interpretation of rule-based models.

# 1. Introduction

Rule-based modeling languages provide a formal means for describing and simulating dynamical phenomena in cellular and molecular biology *(1)*. These languages are typically used to formulate models of cellular regulatory networks *(2)*. The strength and novelty of rule-based models is in their ability to provide concise representations of systems exhibiting combinatorial complexity *(3, 4)*. Instead of reactions or equations, rule-based models are composed of *reaction rules*, which typically define chemical transformations of parts of macromolecules (i.e., sites), as well as the rates at which the transformations occur. Rules are written in a text-based format that is both human and machine readable. Rules are composed of patterns that denote specific molecular moieties. As an example of how the rule-based modeling framework simplifies the model building process, consider a scaffold protein with three binding sites, each of which is specific for a unique kinase protein:

```
Scaf(s1,s2,s3)
K1(s)
K2(s)
K3(s)
```

If each kinase binds a different site on the scaffold protein then the maximum number of scaffold-containing biochemical species that can form is 8. Assuming that the rates of interaction between each kinase and the scaffold are independent of the scaffold's interactions with the other kinases, a model can be defined in terms of only three reaction rules:

```
rule_1: Scaf(s1)+K1(s) <-> Scaf(s1!1).K1(s!1) k1_on,k1_off
rule_2: Scaf(s2)+K2(s) <-> Scaf(s2!1).K2(s!1) k2_on,k2_off
rule_3: Scaf(s3)+K3(s) <-> Scaf(s3!1).K3(s!1) k3_on,k3_off
```

For each rule, note the omission of binding sites that do not participate in the interaction. For example, the first rule only specifies that site `s1` should be unbound (the absence of binding notation associated with a site name indicates the absence of a bond) and so the rule implicitly defines all possible reactions between the scaffold and kinase `K1` regardless of the scaffold's interaction with the other kinases. Rule-based modeling frameworks allow for all possible species to be reached through multiple applications of rules representing interactions in a system to some initial set of biochemical species (termed seed species) while maintaining a relatively simple representation of the entire system.

The above rules are written in the BioNetGen language (BNGL) syntax *(5)*. BNGL was originally developed as a method for automatic generation of reaction networks by determining all possible reactions given a set of seed species and a set of reaction rules. Models written in BNGL can be converted into a system of ordinary differential equations and numerically integrated or converted into a Markov chain describing stochastic chemical kinetics and simulated using a kinetic Monte Carlo (KMC) algorithm *(6)*. Furthermore, network-free KMC methods involving direct application of rules to molecular objects using pattern matching algorithms allow simulation without needing to enumerate the sets of possible reactions and chemical species *(7)*. This flexibility, along with available BNGL-compatible model building and analysis tools *(8–11)*, make rule-based modeling frameworks useful for simulation and analysis of dynamical systems in biology.

While arguably more legible than traditional reaction- or ODE-based representations of reaction networks, reaction rules for some systems can be lengthy and therefore difficult to parse by eye. For example, the mast cell immune response to antigen requires transphosphorylation of the kinase Syk, which occurs when both a catalytically active Syk and a viable substrate Syk are bound to distinct transmembrane receptors that are themselves crosslinked by a multivalent antigen through interaction with receptor-bound IgE antibody.  A rule for Syk transphosphorylation can be difficult to understand simply because the textual representation is lengthy (see Note 1):

```
Syk(SH2!1,kinase~a).FceRI_IgE(gamma!1,Fab!2).Ag(epitope!2,\
   epitope!3).FceRI_IgE(Fab!3,gamma!4).Syk(SH2!4,Tyr~0) -> \
Syk(SH2!1,kinase~a).FceRI_IgE(gamma!1,Fab!2).Ag(epitope!2,\
   epitope!3).FceRI_IgE(Fab!3,gamma!4).Syk(SH2!4,Tyr~P) k_cat
```

However, BNGL is based on a graphical formalism, so we can easily represent such rules visually, making them far easier to read.

There are a few existing tools that facilitate visualization of BioNetGen patterns as graphical objects.  BioNetGen has its own tool, the 'visualize' command for use in the 'actions' block of a plain-text model file (i.e., a BioNetGen input file or BNGL file).  This command can be used to build model summary visualizations in the form of contact maps or regulatory graphs *(12)*.  It can also output a file that includes the graphical representation of all rules in a model, which must then be loaded into a visualization program that can parse GraphML files.  However, the 'visualize' command cannot draw arbitrary BNGL patterns (such as the left-hand side of a rule alone, an observable pattern, or a chemical species definition).  Another useful tool, VCell, employs the BioNetGen engine to construct and simulate rule-based models using a graphical interface *(13)*.  However, it is not possible to rapidly visualize arbitrary patterns or rules simply by copying and pasting a BNGL string into the application.  VCell requires a well-defined model (obtained either by construction of the model through its graphical interface, or by importing a complete BNGL model file) prior to visualization of patterns or rules or export of BNGL strings to a file.

Here, we present a tutorial on how to use a new standalone version of RuleBuilder, which is an application for rapidly visualizing individual patterns and rules written in BNGL. Originally a part of the web-based resource GetBonNie *(14)*, this tool is useful for understanding complex patterns and rules through visualization and constructing BNGL strings de novo via the use of drawing tools. Plain-text BNGL-formatted rules generated by RuleBuilder can be directly inserted into the reaction rules block of a BNGL model file and other patterns may be inserted into BNGL model files as the pattern component of a seed species or observable definition. Ultimately, RuleBuilder is a simple, minimal program for drawing BioNetGen rules and pattern strings in a graph-based format in accordance with the conventions of Faeder *et al*. *(15)*. In the following sections, we will outline what is needed to install and use RuleBuilder to read or write BNGL strings and provide examples illustrating its most notable features.

## 2. Materials

RuleBuilder is written in the Java programming language, and therefore requires a Java Runtime Environment. The source code is available under the BSD-3 license and is hosted at [https://github.com/RuleWorld/RuleBuilder](https://github.com/RuleWorld/RuleBuilder). Compilation of RuleBuilder requires version 8 of the Java Development Kit.

Once compiled, the program is packaged as an executable Java Archive (JAR) and can be run on Unix-like platforms by using the command line to navigate to the directory where the JAR is located and executing the following command:

```
java -jar RuleBuilder.jar
```

It can also be run on Windows or macOS platforms simply by double-clicking on the JAR file.

# 3. Methods

## 3.1. Visualizing a BNGL pattern or reaction rule

One potential use of this tool is to assist in reading existing models or to check BNGL patterns or reaction rules for semantic accuracy while writing a model. In these cases, the textual representation already exists, and can be entered into the input text box of the tool. The string can be pasted or typed directly in the box marked 'BNGL String' and a graphical representation will be generated on the fly. A simple example illustrating how RuleBuilder generates a visualization of a reaction rule can be seen in Figure 1.

RuleBuilder allows manual rearrangement of the graphical elements representing a BNGL pattern or rule. In some cases, patterns may involve symmetric or chain-like polymers (Figure 2A) and a well-organized visualization can facilitate an understanding of the pattern or rule. In these instances, manually arranging the graphical elements can help prevent semantic mistakes (Figure 2B). All that is required for rearranging is to make sure that RuleBuilder is in 'Object Manipulation' mode as specified in the bottom left corner of the RuleBuilder application window. If not in this mode, simply clicking on the arrow icon on the far left of the toolbar will activate this mode.

## 3.2. Graphically building a BNGL pattern or reaction rule

RuleBuilder creates BNGL strings corresponding to user-defined graphical objects. The toolbar at the top of the application window as seen in all figures has a number of icons that toggle various modes of operation (see Note 2). The fifth icon from the left (a gray oval) activates the 'Add Molecule' mode in which additional molecules can be added to the panel. When this mode is active, clicking anywhere in the drawing canvas (taking care not to drag the cursor) adds a new molecule with the default name 'M()' to the canvas. The name can be changed by modifying the string in the text box at the bottom of the application window, or by activating 'Object Manipulation' mode (by clicking on the arrow icon on the top left) and then clicking on the name attached to the graphical representation of the molecule.

Sites can be added to the molecule by choosing from one of the three 'Add Site' modes. These modes' icons are second to fourth from the left and create sites with three distinct bond states: sites where bond state is arbitrary, sites with a bond to an unspecified partner, or sites without a bond. If unbound sites are added, bonds can be added later if necessary using the 'Add Bond' mode. The 'Add Bond' mode is activated by clicking on the sixth icon from the left in the toolbar. After the 'Add Bond' mode is activated, adding a bond requires clicking sequentially on both sites that are intended to be represented as bound together. Sites may have internal states. Internal states are added or changed in a manner similar to changing molecule or site names; in accordance with BNGL syntax, a '~' character should be added to separate the site name from the state label (see Note 3).

To build a rule, specific syntax denoting which molecules are reactants and products is required. To specify molecules that are in separate chemical species a 'plus' sign must be placed between the patterns (fifth icon from the right). Furthermore, reactants should be placed to the left of an arrow (fourth and third icon from the right) and products should be placed to the right (see Note 4). Rate law information is automatically incorporated into the constructed BNGL string depending on the selected arrow (unidirectional or bidirectional) and the string denoting the rule's rate law can be changed as one would change molecule or site names (see Note 5).

BNGL rules involve a mapping of sites on the left-hand side to sites on the right-hand side. This mapping is ordinarily inferred by the BioNetGen software, but it may also be defined manually by a user. RuleBuilder thus provides an 'Add Mapping' mode (seventh icon from the left) to allow construction of a mapping, which is similar in operation to the 'Add Bond' mode when creating bonds. Note that sites can only be mapped from reactants to products, meaning that an arrow operator must already be present in the rule prior to using the graphical interface's mapping feature.

## 3.3. A rule-building example

In this section, we will walk through building a rule using the graphical interface, which will generate a BNGL string that can then be copied into a BNGL file. The rule will characterize a reversible interaction between a dimer (scaffold bound to a kinase) and a monomer. The product is a trimer (scaffold bound to two kinases on distinct binding sites). The BNGL string encoding the rule is as follows:

```
Scaf(s1!1,s2).K1(s!1) + K2(s2) <-> \
Scaf(s1!2,s2!3).K1(s!2).K2(s2!3) k_on,k_off
```

It should be noted that at any time during the drawing process, unwanted objects can be removed by simply clicking on the arrow icon (switching to 'Object Manipulation' mode), then clicking on the unwanted object, and finally clicking the trash icon to delete the selected object.

Furthermore, the layout of graphical objects can be adjusted at any time throughout the drawing process, and individual molecule objects can be resized by clicking on them in 'Object Manipulation' mode and dragging the boxes surrounding the object.

First, we will construct the left-hand side of the rule given above. We create three new molecules using 'Add Molecule' mode by clicking on the gray ellipse icon and then clicking three times in the blank drawing canvas. You should see three ellipses each labeled 'M' and a string in the bottom text box reading:

```
M().M().M()
```

Note that RuleBuilder takes the molecules to be joined together in a complex (see Note 6). Molecules that are in distinct chemical species should be separated by the '+' sign. Next, we will change the names of the molecules simply by modifying the BNGL string in the text box (Figure 3A) to read as follows:

```
Scaf().K1().K2()
```

We will now add the necessary sites and the plus operator to complete the rule's left-hand side as described below. In our example, we have a scaffold bound to a kinase interacting with a second kinase. Therefore, the scaffold needs to have one free site and one site bound to one kinase. To achieve this, we first activate the 'Add Site (unbound)' mode by clicking on the icon with the full white circle and then we can add two sites to the scaffold and one site each to the kinases by clicking twice on the scaffold object and once on each kinase object. After renaming the sites. we activate the 'Add Bond' mode and connect one site of the scaffold to the site in kinase `K1`. Finally, we insert the plus sign by clicking on the plus icon and clicking between the dimer and monomer patterns (Figure 3B).

From here, we can define a reversible reaction rule by clicking on the double-arrow icon and then clicking on the drawing canvas to the right of the second kinase molecule, K2. The application will automatically add rate constants (forward and reverse, in this order), which can be renamed if desired. We then construct the right-hand side of the rule as we did the left but we add a bond between the interacting kinase and scaffold to indicate the product of the interaction. We omit the plus operator that is present on the left-hand side, because the scaffold and kinase are now members of the same chemical species (Figure 3C).

# 4. Notes

1. The '\' character at the end of a line denotes line-continuation in BNGL.
2. From left to right, the icons on the toolbar correspond to the following modes:
    a. Object Manipulation Mode
    b. Add Site (arbitrary bond state)
    c. Add Site (bound, unspecified binding partner)
    d. Add Site (unbound)
    e. Add Molecule
    f. Add Bond
    g. Add Mapping
    h. Add Plus Sign
    i. Add Arrow Separator
    j. Add Double-Arrow Separator

   The final two icons are a trash can and a disk icon. The trash can icon enables deletion of objects from the graph (provided that an object is selected via 'Object Manipulation' mode). The disk icon enables saving the current image to file in multiple formats, including the GIF, PNG, and JPEG formats.

3. Sites can only have one internal state, meaning that only one '~' character is allowed following a site's name.
4. RuleBuilder does not fully parse BNGL syntax. As a result, strings denoting a rule's rate law information should be restricted to alphanumeric symbols to avoid erroneous parsing of the rule.
5. The double-arrow operator is syntactic sugar allowing two unidirectional rules to be written on one line as a bidirectional (i.e., reversible) rule. One unidirectional rule is defined by a left-to-right reading and the other is defined by a right-to-left reading. Rules defined with the double-arrow operator require two rate law definitions separated by a comma.
6. Molecules that do not have sites but are connected with the dot (.) operator are part of the same complex. This is not shown in RuleBuilder, as bonds are only drawn between sites. Molecules shown in the RuleBuilder GUI are assumed to be connected unless the '+' operator is between them.

## 5. Acknowledgement

This work was supported by NIH/NIGMS grant R01GM111510. RS also acknowledges support from the Center for Nonlinear Studies, which is funded by the Laboratory Directed Research and Development program at Los Alamos National Laboratory, which is operated for the National Nuclear Security Administration of the U.S. Department of Energy under contract DE-AC52-06NA25396.

# Figure captions

1. A visual representation of a simple rule, generated by a BNGL string based on the scaffold-kinase example presented in the text.

2. (A) The default layout for representation of a multivalent antigen-bivalent antibody aggregate. The full BNGL string (not fully visible in the text box) is:

   `IgE(Fab,Fab!0).IgE(Fab!1,Fab!2).Ag(s,s!1,s!0).IgE(Fab!3,Fab).Ag(s!2,s!4,s!3).IgE(Fab!4,Fab)`

   (B) A manually arranged layout for the same aggregate visualized in panel A.

3. (A) Create a three-molecule pattern using the 'Add Molecules' mode to begin construction of a rule. (B) Sites and bonds are added to the pattern, providing context to define the reactants to which the rule applies. (C) To complete the rule, products are added to the right-hand side of the rule to define the rule's transformation.

RuleBuilder

# Figure 1

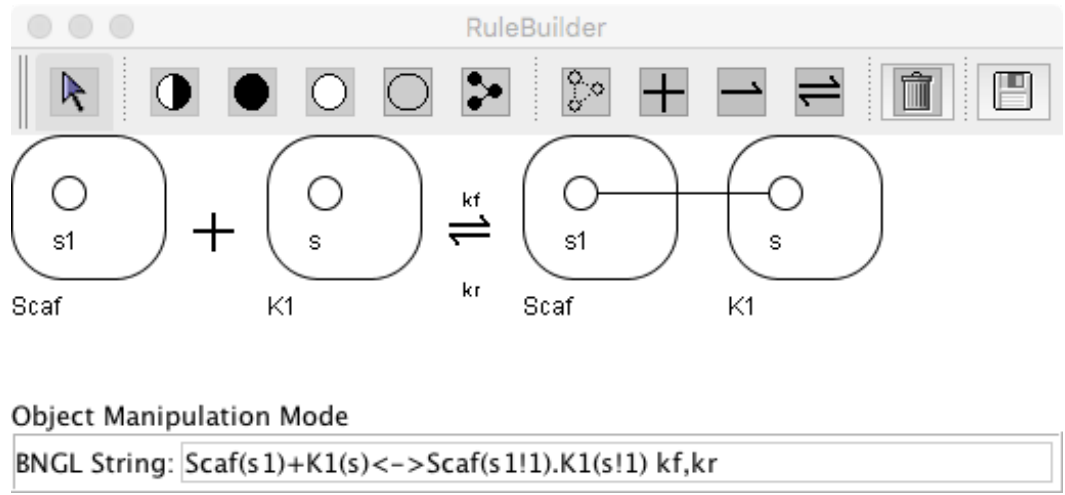

# Figure 2

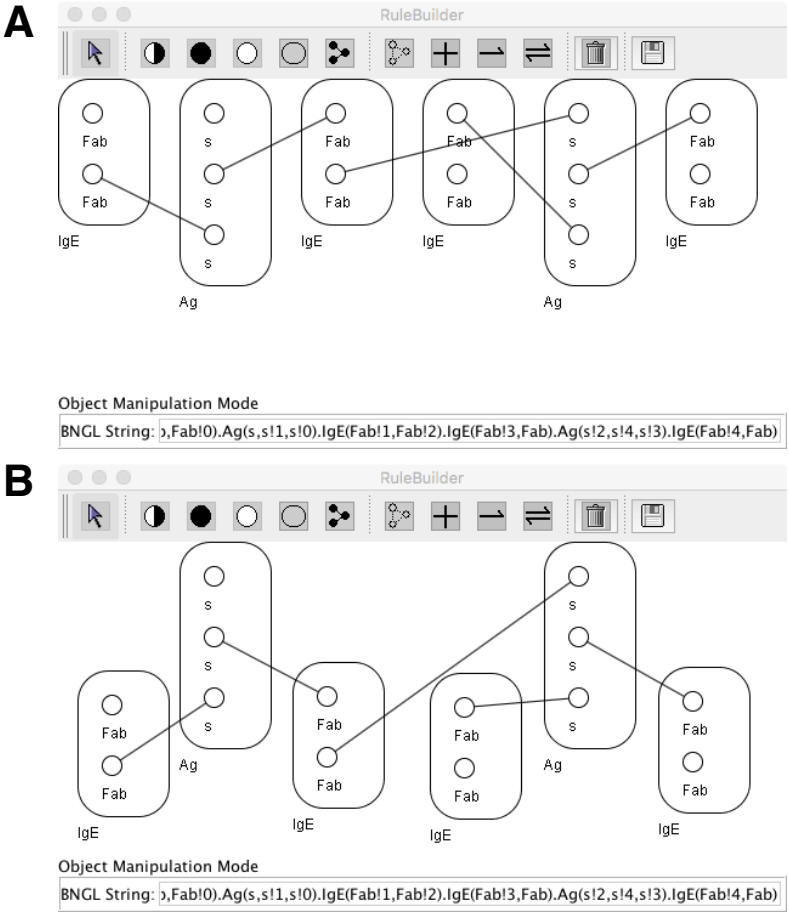

RuleBuilder

# Figure 3

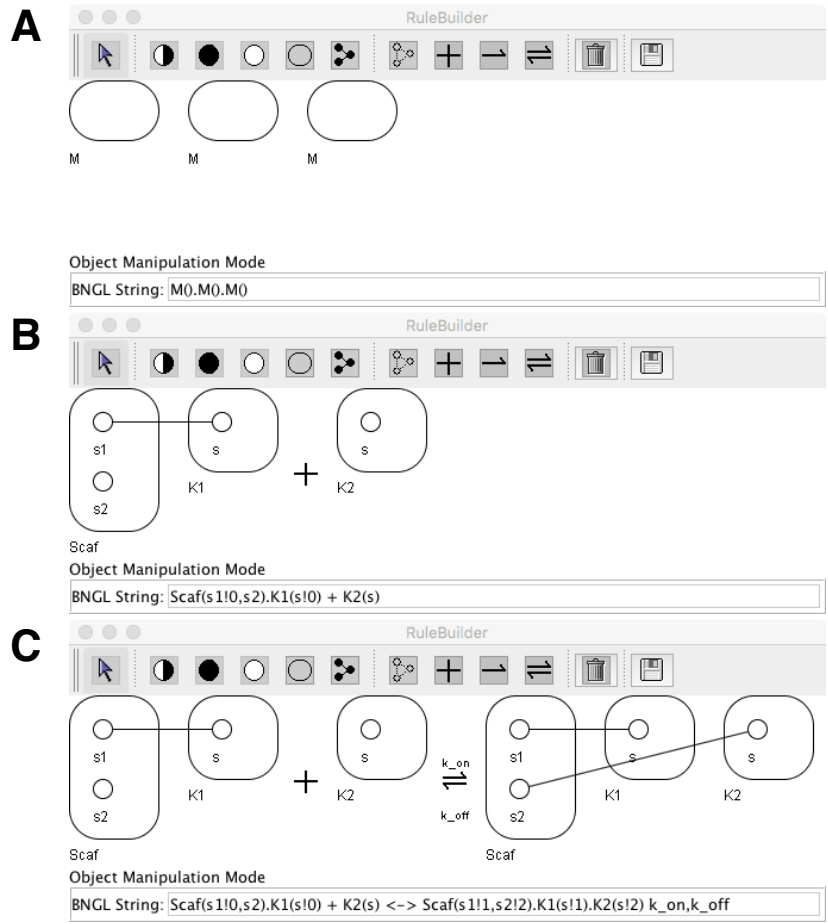